\documentstyle[aps,preprint]{revtex}
\begin{document}
\draft
\title{\bf Combinatorial identities for binary necklaces from exact
ray-splitting trace formulae}
\author
{R. Bl\"umel and Yu. Dabaghian}
\address
{Department of Physics, Wesleyan University,
Middletown, CT 06459-0155}
\date{\today}
\maketitle
\begin{abstract}
Based on an exact trace formula for a one-dimensional ray-splitting system,
we derive novel combinatorial identities for cyclic binary
sequences (P\'olya necklaces).
\end{abstract}
\section{Introduction}
Wave propagation in systems with sharp interfaces is a
fundamental problem in the natural sciences and engineering.
Well-known examples include light waves impinging on a water-air
interface or sound waves propagating in layered media \cite{Bre}.
All these systems have one feature in common: the splitting
of the incident wave into reflected and transmitted components.
In the geometrical optics limit of small wave lengths, incident,
reflected and transmitted waves are described by rays.
The rays are split at the interface; hence the name
``ray-splitting systems'' \cite{Couch} for the whole class
of wave systems with sharp interfaces. Ray-splitting systems
have recently attracted attention in the context of
acoustic, quantum and electromagnetic wave chaos
\cite{Couch,QRS,Kill,Prange,Annals,SKB,Bauch,Fest,us}.
It was shown that the mere presence of a ray-splitting boundary
can drive an otherwise regular system into chaos \cite{Couch,QRS,Kill}.
It was also shown that ray-splitting systems produce
corrections to the Weyl formula \cite{W1,W2}
for the average density of states
that can be computed analytically \cite{Prange,Annals}.
The most conspicuous consequence of ray splitting is the
existence of non-Newtonian periodic orbits in a ray-splitting
system that contribute substantially to the fluctuating
part of the level density \cite{QRS,Kill,SKB,Bauch,Fest}. The existence
of non-Newtonian orbits in a
dielectric-loaded Bunimovich ray-splitting stadium
was demonstrated experimentally \cite{SKB,Bauch,Fest}.
In addition it
has been shown recently that exact trace formulae
exist for
a class of one-dimensional ray-splitting systems
\cite{Fest,us}. These formulae were derived using
results from quantum graph theory \cite{QGT1,QGT2}.
Considering the two-point correlation function of
the spectra of special quantum graphs,
Schanz and Smilansky
were able to derive novel combinatorial identities \cite{combi}.
This was possible by deriving the two-point correlation function
in two independent ways, (i) directly using input from the
quantum spectrum and (ii) using the exactness of the trace formula.
Motivated by
the methods of Kottos, Schanz and
Smilansky \cite{QGT1,QGT2,combi}
we show that
novel combinatorial identities for binary
P\'olya necklaces \cite{necklace,intro} are obtained directly
by comparing the spectral
density of analytically solvable quantum graphs with their
exact periodic orbit expansion.

The plan of this paper is as follows: in section II we present
our model system, a one-dimensional ray-splitting system, whose
spectrum can be obtained analytically. We present an exact
periodic-orbit expansion of its level density
expressed as a generalized Fourier sum over binary P\'olya
necklaces.
In section III we outline our method for
obtaining exact combinatorial identities for binary
necklaces derived from
analytically solvable cases of the one-dimensional
ray-splitting system. In section IV we present two worked
examples that yield two infinite sets of combinatorial
identities. In section V we discuss our results and conclude the
paper.

\section{Spectrum}
Denote by $E_{n}$ the spectrum of the
one-dimensional scaled Schr\"odinger equation
\begin{equation}
-\psi''(x)+V_{\lambda}(E,x)\psi(x)=E\psi(x),
\label{schre}
\end{equation}
where
\begin{equation}
V_{\lambda}(E,x) =\cases{0, &for $0<x\leq a$, \cr
\lambda E, &for $a<x<1$,\cr
        \infty, &for $x\not \in (0,1)$ \cr}
\label{poten}
\end{equation}
and $\psi(x)\equiv 0$ for $x\not\in (0,1)$. Scaling
potentials of the form (\ref{poten}) arise naturally
in many ray-splitting systems, for instance in
dielectric-loaded cavities \cite{Prange,SKB,Bauch,Fest}.
In this paper we concentrate on the case
$E> V_{\lambda}(E,x)$ for all $x$. Then, without
restriction of generality,
the scaling constant $\lambda$ can be assumed to
satisfy $0\leq \lambda < 1$.
Define $E=k^{2}$, then the spectrum of (\ref{schre}) is determined by
\begin{equation}
\sin\left[k(\sigma_{L}+\sigma_{R})\right] -
r\sin\left[k(\sigma_{L}-\sigma_{R})\right]=0,
\label{eqn}
\end{equation}
where $\sigma_{L}=a$,
$\sigma_{R}=\eta(1-a)$, $\eta=\sqrt{1-\lambda}$, and
$r=(1-\eta)/(1+\eta)$ is the reflection coefficient.
For the derivations below it
is useful to define the transmission coefficient
$t=\sqrt{1-r^2}$. Therefore, the reflection and
transmission coefficients satisfy the relation
\begin{equation}
r^2+t^2=1.
\label{reftran}
\end{equation}
Since in this paper we focus on the case $0\leq \lambda<1$,
both $r$ and $t$ are real and positive and
range between 0 and 1.
Possible quantum phases incurred at reflection or transmission
events
are treated explicitly and separately (see, e.g.,
(\ref{gutzw}) below). They
are not included in $r$ or $t$.

We now discuss an alternative method of solving the quantum
dynamics in the potential (\ref{poten}). This method is based
on coding the periodic orbits with the help
of symbol strings.
We denote a bounce
off $x=0$ by the letter $\cal{L}$ and a bounce off
$x=1$ by the letter $\cal{R}$. Words formed with these two
letters code for periodic orbits in (\ref{poten}).
The word $\cal{L}$, for instance, codes for the non-Newtonian
orbit that bounces between $x=0$ and $x=a$ (above-barrier
reflection orbit). The word
$\cal{LR}$ codes for the (Newtonian) orbit that bounces
between $x=0$ and $x=1$. Since a periodic orbit represented
by the word $w$ cycles
through the letters of $w$ without a well-defined beginning or
end, two words $w$ and $w'$ are equivalent in our context,
and code for the same periodic orbit, if
they are of the same length (i.e. they consist of the
same number of symbols) and
their respective symbol sequences are identical up to
cyclic permutations.
Sequences of objects that are identical up
to cyclic permutations are called (P\'olya) necklaces
\cite{necklace,intro}.
If the number of objects they consist of is two, they are
called binary necklaces. Apparently therefore,
the periodic orbits of (\ref{poten}) can be coded with
the help of binary necklaces over the symbols
$\cal{L}$ and $\cal{R}$.
It is remarkable that for (\ref{poten}) every
Newtonian or non-Newtonian
periodic orbit can be mapped
one-to-one onto a binary necklace. In other words,
``pruning'' is not necessary for
the binary necklaces relevant for
(\ref{poten}).
Every binary necklace defines a possible
periodic orbit of (\ref{poten}) and vice versa.

Given two letters, for instance $\cal{L}$ and $\cal{R}$, we
can form $2^{\ell}$ words of length $\ell$. But, in general,
many of these words will be cyclically equivalent, and
correspond to the same necklace. So, how many necklaces of
length $\ell$ are there? This question is answered by
the following formula. There are exactly \cite{intro}
\begin{equation}
N(\ell) = {1\over \ell}\, \sum_{n|\ell}\,
\phi(n)\, 2^{\ell/n}
\label{ncount}
\end{equation}
binary necklaces of length $\ell$,
where the symbol ``$n|\ell$'' denotes ``$n$ is a divisor
of $\ell$'', and
$\phi(n)$ is Euler's totient function defined as the
number of positive integers smaller than $n$ and relatively
prime to $n$ with $\phi(1)=1$ as a useful convention.
Thus the first four totients are given by $\phi(1)=1$,
$\phi(2)=1$, $\phi(3)=2$ and $\phi(4)=2$.
We illustrate (\ref{ncount}) with two examples
for $\ell=1$ and $\ell=2$.
There are two necklaces for $\ell=1$,
$\cal{L}$ and $\cal{R}$. Applying (\ref{ncount}) to this
problem, we verify $N(1)=\phi(1)\times 2=2$.
There are three necklaces of length 2, $\cal{LL}$,
$\cal{LR}$ and $\cal{RR}$; again verified by (\ref{ncount}),
$N(2)=[\phi(1)\times 4 + \phi(2)\times 2]/2=3$.

Given a binary necklace $w$, we
define the following integer-valued functions
on $w$: $n_{\cal{R}}(w)$ counts the number of $\cal{R}$s in $w$,
$n_{\cal{L}}(w)$ counts the number of $\cal{L}$s,
$n(w)=n_{\cal L}(w)+n_{\cal{R}}(w)$,
$\chi(w)$ is the sum of $n(w)$ and the
number of $\cal{R}$-pairs in $w$,
$\alpha(w)$ counts all occurrences of
$\cal{R}$-pairs or $\cal{L}$-pairs,
$\beta(w)$ counts all occurrences of
$\cal{RL}$ or $\cal{LR}$ and $\gamma(w)$
is defined as
$\gamma(w)=2n_{\cal L}(w)+n_{\cal{R}}(w)$.
Note that the counting of
$\cal{R}$-pairs, $\cal{L}$-pairs, $\cal{LR}$- or
$\cal{RL}$-combinations is to be understood cyclically, i.e.,
for example, $\alpha({\cal{R}})=1$ and $\beta({\cal{LR}})=2$.
Next we define the set $W_{p}$ of prime necklaces
as the ones that cannot be written as a periodic concatenation of
substrings.
As shown recently \cite{us},
there exists an exact periodic orbit expansion
for the spectral density of (\ref{schre})
in terms of prime binary necklaces,
\begin{equation}
\rho(k)=\bar\rho +
{1\over 2\pi}\sum_{w\in W_{p}}S_{w}
\sum_{\nu=-\infty\atop\nu\neq 0}^{\infty}
\left[(-1)^{\chi(w)}r^{\alpha(w)}t^{\beta(w)}
\right]^{\mid\nu\mid}\,e^{i\nu S_{w}k}, \ \ \ k>0,
\label{gutzw}
\end{equation}
where
\begin{equation}
S_{w}=2[n_{\cal{R}}(w)\sigma_R+
               n_{\cal{L}}(w)\sigma_L]
\label{act}
\end{equation}
is the action of the primitive periodic orbit coded by the
prime binary necklace $w$
and $\bar\rho=(\sigma_L+\sigma_R)/\pi$ is the average level density.

\section{Method}
For special values of the parameters of the potential
well (\ref{poten}) it is
possible to solve (\ref{eqn}) analytically, thus obtaining
directly the density of states $\rho(k)$. Equating the explicit
expression for $\rho(k)$ with the necklace expansion (\ref{gutzw}), one
obtains combinatorial identities for binary necklaces.
An illustrative example is the case $a=1$, for which the
spectral density of (\ref{schre}) is given by
\begin{equation}
\rho(k)=\sum_{m=-\infty}^{\infty}\delta \left(k-\pi m\right).
\label{nostep}
\end{equation}
In this case there exists only one primitive
necklace, ${\cal LR}$, and the necklace
expansion of (\ref{gutzw}) yields
\begin{equation}
\rho(k)=\frac{1}{\pi}\sum_{\nu =-\infty}^{\infty}e^{2i\nu k}.
\label{nostep1}
\end{equation}
Equating (\ref{nostep}) and (\ref{nostep1})
yields the well-known Poisson
formula.

A comment is in order here. For every finite $a$ the
necklace expansion (\ref{gutzw}) involves an infinite sum
over prime periodic necklaces. At $a=1$ this sum collapses
to a single term. One may ask the question how this singular
limit arises. The answer is the following. For every finite
$a$ the actions of the right-hand lobes of the
periodic orbits of (\ref{gutzw}) is finite. At $a=1$, these
actions are zero. Consequently, all necklaces that represent
{\it different} prime periodic orbits for $a\neq 1$ become
{\it repetitions} of the Newtonian
periodic orbit $\cal{LR}$ at $a=1$.
This is the reason
for the existence of only
a single prime periodic orbit ($\cal{LR}$) at $a=1$.

Apart from trivial and well-known
identities such as (\ref{nostep}) and
(\ref{nostep1}) above, (\ref{gutzw}) is a rich source of new and
nontrivial combinatorial identities for binary necklaces.
Specific examples are discussed in Section III. Here we outline
the general method.

Equation (\ref{eqn}) can be written as
\begin{equation}
\sin(\omega_{1}k) - r\sin(\omega_{2}k)=0,
\label{eqn2}
\end{equation}
where $\omega_{1}=\sigma_{L}+\sigma_{R}$ and
$\omega_{2}=\sigma_{L}-\sigma_{R}$.
A negative $\omega_2$
corresponds to a mirror reflection of
$V_{\lambda}(E,x)$ with respect to $x=1/2$.
Thus, because of $\sigma_{L}$,$\sigma_{R}\geq
0$, and without loss of generality,
we may
assume $\omega_{1}\geq \omega_{2} \geq 0$.

In case $\omega_{1}$ and $ \omega_{2}$ are rationally related,
i.e.
$\omega_{1}/\omega_{2}=p/q$, $p\geq q\in {\bf N}$
and $p,q$ relatively prime, (\ref{eqn2}) is
reduced to the algebraic equation

\begin{equation}
\sin(p\omega k) - r\sin(q\omega k)=0,
\label{eqn3}
\end{equation}
where $\omega_{1}=p\omega$ and $\omega_{2}=q\omega$. Using the formula
\begin{equation}
\sin(nx)=\sin(x)\, U_{n-1}(\cos(x)),
\end{equation}
where $U_{n-1}(x)$ is the
Chebyshev polynomial of the second kind, one obtains
\begin{equation}
\sin (\omega k)\left[U_{p-1}(\cos\omega k)-
rU_{q-1}(\cos\omega k)\right]=0.
\label{eqn4}
\end{equation}
It follows immediately from (\ref{eqn4})
that in the case of rationally
related $\omega_{1}$ and $\omega_{2}$
there always exists a sequence of
roots $k^{(0)}_{n}=\pi n/\omega$.
The remaining roots are determined by
\begin{equation}
U_{p-1}(x)- rU_{q-1}(x)=0,
\label{u}
\end{equation}
where $x=\cos(\omega k)$.
Since every root $x_{j}$ of (\ref{u})
gives rise to a periodic sequence of
eigenvalues, $\cos(\omega k^{(j)}_n)=x_{j}$, $j=1,2,...,p-1$,
together with the sequence $k_n^{(0)}$
the spectrum of
(\ref{eqn2}) consists of $p$ (possibly degenerate)
periodic sequences of roots.
Whenever
(\ref{u}) can be solved analytically,
the density of states
\begin{equation}
\rho(k) = \sum_{j=0}^{p-1}\, \sum_{n=-\infty}^{\infty}\,
\delta\left( k-k_n^{(j)}\right)
\label{explicit}
\end{equation}
is known explicitly and together with (\ref{gutzw})
leads to a host of combinatorial identities
for binary necklaces.
Two examples are presented in the following
section.

\section {Combinatorial identities}
\par\noindent
{\bf Example 1}.
For $\sigma_{L}=\sigma_{R}$
equation (\ref{eqn}) becomes
\begin{equation}
\sin(2ka)=0,
\label{first}
\end{equation}
with the solutions $k_{n}=\pi n/(2 a)$. Note that there is no
$r$-dependence in (\ref{first}). The density of states is given by
\begin{equation}
\rho(k)=
\sum_{n=-\infty}^{\infty}\delta \left(k-\frac{\pi n}{2a}\right)=
\frac{2a}{\pi}\sum_{m=-\infty}^{\infty}e^{4imak}.
\label{onezero}
\end{equation}
According to (\ref{act}) and
due to
$\sigma_{L}=\sigma_{R}$, $S_w$ depends only on the binary
length of $w$ and is given by $S_w=2an(w)$. Thus
the sum (\ref{gutzw}) can be
written as
\begin{equation}
\rho(k) = {2a\over \pi} + {a\over\pi}
\sum_{m=-\infty\atop m\neq 0}^{\infty}\
\sum_{w\cdot \nu \in W_{\mid m\mid }}
n(w) \, \left[(-1)^{\chi(w)}
r^{\alpha(w)}
t^{\beta(w)}
\right]^{\nu} \, e^{2imka},
\label{gutzw1}
\end{equation}
where $W_{n}$ denotes the set of all length-$n$ binary necklaces
{\rm w}, $w$ is the shortest primitive code-piece in {\rm w}
and $\nu$ is the number of its repetitions in {\rm w}.
Comparing the series (\ref{onezero}) and (\ref{gutzw1}),
we see that odd-length and even-length
binary necklaces satisfy the sum rules
\begin{equation}
\sum_{w\cdot \nu \in W_{2m-1}} n(w)
\left[(-1)^{\chi(w)}r^{\alpha(w)}t^{\beta(w)}\right]^{\nu}=0,
\ \ \ m=1,2,...\,
\label{sum0}
\end{equation}
and
\begin{equation}
{1\over 2}\, \sum_{w\cdot \nu \in W_{2m}} n(w)
\left[(-1)^{\chi(w)}r^{\alpha(w)}t^{\beta(w)}\right]^{\nu}=1,
\ \ \ m=1,2,...\, .
\label{sum1}
\end{equation}
At first glance it may seem surprising that (\ref{sum1})
is a
constant for all $r$. The solution lies in the relation
(\ref{reftran}). When properly ordered according to powers
of $r$ and $t$, it turns out that (i)
individual terms in (\ref{sum1}) are of the form
$r^{2p}t^{2q}$, where $p+q=m$ and (ii)
the coefficients in front of the term $r^{2p}t^{2q}$ in
(\ref{sum1}) turn out to be binomial coefficients.
Thus, for given $m$, the left-hand side of (\ref{sum1})
reduces to $(r^2+t^2)^m$, which, according to (\ref{reftran})
is equal to 1 for any
choice of $r$. This explains why the
seemingly variable
left-hand side of (\ref{sum1}) is nevertheless a constant.
Thus we obtain
from (\ref{sum1}) the following
infinite set of combinatorial
identities for even-length binary necklaces
\begin{equation}
{1\over 2}\, \sum_{w\cdot \nu \in W_{2m}} n(w)
(-1)^{\nu\cdot\chi(w)}\,
\delta_{\nu\cdot\alpha(w)/2,s}\ =\
\left(\matrix{m\cr s\cr}\right),
\ \ \ s=0,...,m,\ \ m=1,2,...\, ,
\label{eq19}
\end{equation}
where $\delta_{i,j}$ is the Kronecker symbol.

In order to illustrate (\ref{eq19}) let us first
focus on the case $m=1$.
According to (\ref{ncount})
there are exactly three
cyclically non-equivalent
necklaces of
binary length 2 given by
${\rm w}_1=\cal{LL}$,
${\rm w}_2=\cal{LR}$,
${\rm w}_3=\cal{RR}$.
The necklaces ${\rm w}_1$ and ${\rm w}_3$ are not primitive.
The necklace
${\rm w}_1$ is a two-fold repetition of the primitive necklace
$w_1=\cal{L}$. Thus
$\nu_1=2$. An analogous consideration for
${\rm w}_3$ yields
$w_3=\cal{R}$ and
$\nu_3=2$.
The necklace ${\rm w}_2$ is primitive. Therefore
$w_2={\rm w}_2=\cal{LR}$ and
$\nu_2=1$.
The three necklaces ${\rm w}_j$, $j=1,2,3$,
are listed in Table~I. Also listed are their
primitives $w_j$, the repetition indices $\nu_j$,
and the values of the functions $n(w_j)$,
$\alpha(w_j)$, $\beta(w_j)$, $\gamma(w_j)$
and $\chi(w_j)$.

We are now ready to check (\ref{eq19}).
For $m=1$ we have two choices for $s$: $s=0$ and $s=1$.
For $s=0$ we have to scan the three words $w_j$, $j=1,2,3$,
for $\nu_j \alpha(w_j)/2=s=0$. According to the entries in
Table~I, only ${\rm w}_2$
qualifies and the sum on the left-hand side of
(\ref{eq19}) reduces to the single term
\begin{equation}
{1\over 2} n(w_2)(-1)^{\nu_2\chi(w_2)} = 1 =
\left(\matrix{1\cr 0\cr}\right).
\end{equation}
This shows that (\ref{eq19}) is indeed true for the simplest
case $m=1$, $s=0$. For the case $m=1$, $s=1$ we have to
check Table~I for occurrences with $\nu_j\alpha(w_j)/2=1$.
This is fulfilled for the necklaces ${\rm w}_1$ and ${\rm w}_3$.
We obtain
\begin{equation}
{1\over 2}\left[n(w_1)(-1)^{\nu_1\chi(w_1)} +
n(w_3)(-1)^{\nu_3\chi(w_3)}  \right] = 1 =
\left(\matrix{1\cr 1\cr}\right).
\end{equation}
This shows that (\ref{eq19}) also works for $m=1$, $s=1$.

Testing (\ref{eq19}) for $m=2$ involves finding all
non-equivalent necklaces of binary length 4.
According to (\ref{ncount})
there are exactly six.
All six necklaces are listed in Table~II
together with their properties.
For $m=2$ we have three possibilities for $s$: $s=0,1,2$.
For $s=0$ we have to check Table~II for necklaces that fulfill
$\nu_j\alpha(w_j)/2=0$. Only ${\rm w}_4$ qualifies. We obtain
$n(w_4)/2=1$, which equals
$\left(\matrix{2\cr 0\cr}\right)$,
the binomial coefficient on the
right-hand side of (\ref{eq19}). For $s=1$ we
have to check for $\nu_j\alpha(w_j)/2=1$. We find three
candidates: ${\rm w}_2$, ${\rm w}_3$ and ${\rm w}_5$. This time
we have to be careful when summing the three terms on the
left-hand side of (\ref{eq19}),
since
$\nu_3\chi(w_3)=5$. Therefore the second term
in (\ref{eq19})
contributes with a
minus sign. We obtain $[n(w_2)-n(w_3)+n(w_5)]/2 = 2$, which
equals
$\left(\matrix{2\cr 1\cr}\right)$,
the corresponding binomial coefficient on the
right-hand side of (\ref{eq19}).
Two necklaces, ${\rm w}_1$ and ${\rm w}_6$,
contribute in the case $s=2$ and again satisfy (\ref{eq19}).

\noindent
{\bf Example 2}.
Suppose now that
$\sigma_{L}=2\sigma_{R}$. In this case
(\ref{eqn}) becomes
\begin{equation}
\sin(ka/2)\,
\left[4\cos^{2}(ka/2)-r-1\right]=0.
\end{equation}
This equation has three sets of solutions
\begin{equation}
k_n^{(j)} ={2j\over a}\arccos(\varphi)+{2\pi n\over a},\ \ \
j=-1,0,1,
\label{roots2}
\end{equation}
where $\varphi=\sqrt{1+r}/2$.
The density of states is
\begin{eqnarray}
\rho(k)=\sum_{j=-1}^1\,
\sum_{m=-\infty}^{\infty}\delta \left(k+{2j\over a}
\arccos(\varphi)-\frac{2\pi m}{a}\right) =
\cr
\frac{a}{2\pi}\sum_{j=-1}^1 \sum_{n=-\infty}^{\infty}
e^{in[ak+ 2j\arccos(\varphi)]}
=\frac{a}{2\pi} \sum_{n =-\infty}^{\infty}e^{inak}
\left[ 2T_{2n}(\varphi)+1\right],
\label{onetwo}
\end{eqnarray}
where $T_{n}(x)\equiv \cos (n \arccos x)$ are the
Chebyshev polynomials of
the first kind \cite{Atlas}.
Equating (\ref{onetwo}) order-by-order with the
necklace expansion
(\ref{gutzw})
we obtain the sum rules
\begin{eqnarray}
\sum_{w\in W_p}\sum_{\nu=1}^{\infty}\gamma(w)\left[
(-1)^{\chi(w)} r^{\alpha(w)} (1-r^{2})^{\beta(w)/2}\right]^{\nu}\,
\delta_{\nu\gamma(w),m}=
\cr
1+\sum_{j=0}^{m}
\frac{2m(-1)^{j}}{2m-j}
\pmatrix{2m-j \cr j} (1+r)^{m-j},\ \ m=1,2,...\, .
\label{sum2}
\end{eqnarray}
We used formula 22:6:1 of ref. \cite{Atlas} for the
Chebyshev polynomials in (\ref{onetwo}).

Ordering (\ref{sum2}) according to powers of $r$, (\ref{sum2})
can be reformulated
as a combinatorial theorem on
the set of binary necklaces,
in which $\cal{L}$ beads weigh twice as much as $\cal{R}$ beads:
$$
\sum_{w\in W_p,C}\gamma(w)
(-1)^{[2m\chi(w)+s\gamma(w)-m\alpha(w)]/[2\gamma(w)]}\,
\left(\matrix{{m\beta(w)\over 2\gamma(w)}\cr
              {s\gamma(w)-m\alpha(w)\over 2\gamma(w)}\cr}\right)
\ = \
$$

\begin{equation}
\delta_{s,0} +
\sum_{j=0}^{m-s}{2m(-1)^j\over 2m-j}\,
\left(\matrix{2m-j\cr j\cr} \right)
\, \left(\matrix{m-j\cr s\cr} \right) ,
\ \ \ s=0,1,...m,\ \ \ m=1,2,...\, .
\label{pol}
\end{equation}
The condition $C$ in the sum (\ref{pol}) is
$C=\gamma(w)|m\, \wedge\, s-m\alpha(w)/\gamma(w)$ even.
The sum on the left-hand side of (\ref{pol}) may be empty.
In this case the sum is defined to be zero.

Let us check (\ref{pol}) with the help of a few examples.
First we focus on the case $m=1$, $s=0$. In order to fulfill
the first part of the condition $C$ in (\ref{pol})
we need $\gamma=1$.
This, in turn, requires to find a necklace with
$n_{\cal{L}}=0$ and
$n_{\cal{R}}=1$. There is just one such necklace, namely
$\cal{R}$. But it does not fulfill the second
part of $C$.
Therefore, the sum on the left-hand side of (\ref{pol})
is empty, and the left-hand side is zero. The right-hand
side adds up to $1+1-2=0$ and confirms (\ref{pol})
for this special case. For $s=1$ we find again that
$\cal{R}$ is the only choice for $w$. But this time the
second part of $C$ is fulfilled and the left-hand side
of (\ref{pol}) is
\begin{equation}
\gamma({\cal{R}})(-1)^{[2\chi({\cal{R}})+\gamma({\cal{R}})-
\alpha({\cal{R}})]/[2\gamma({\cal{R}})]}
\left(
\matrix{\beta({\cal{R}})/[2\gamma({\cal{R}})]\cr
[\gamma({\cal{R}})-\alpha({\cal{R}})]/[2\gamma({\cal{R}})]\cr} \right)
= 1.
\end{equation}
We used $\alpha({\cal{R}})=1$,
$\beta({\cal{R}})=0$,
$\gamma({\cal{R}})=1$ and
$\chi({\cal{R}})=2$.
For $m=1$, $s=1$
the right-hand side of (\ref{pol}) consists of just one term,
which turns out to be 1 as well. Thus we checked that
(\ref{pol}) works
for $m=1$. With the help of Tables~I and II,
other special cases may be checked as well.

\bigskip

\section{Discussion and Conclusions}
The work presented here is closely related to the
theory of quantum graphs \cite{QGT1,QGT2,combi}.
While the quantum graphs considered by Kottos, Schanz and
Smilansky \cite{QGT1,QGT2,combi}
correspond to the case of zero potential
on the bonds and delta potentials on the vertices,
the step
potentials considered in this paper correspond to
constant potentials on the bonds and
potential steps at the vertices. Thus, although the
methods employed in this paper are essentially
those used previously by Kottos, Schanz and Smilansky,
we obtain a different class of
combinatorial identities that apply to
cyclic binary codes (P\'olya necklaces).
Another difference concerns the derivation of
identities. While Kottos, Schanz and Smilansky use a route
that involves two-point correlation functions, we show that
novel
combinatorial identities can be obtained directly from
the periodic orbit expansions of explicitly solvable cases.
These minor differences notwithstanding the central idea
for generating entirely new classes of combinatorial
identities is the same: Combinatorial identities can be obtained
whenever a quantum system admits of (i) an explicit analytical
solution and (ii) an exact periodic orbit expansion.

In addition to the two examples presented above, there exist
many other cases in which (\ref{u}) can be reduced to a low-order
polynomial that can be solved by elementary means.
Examples are the cases $p=3$, $q=2$ or
$p=5$, $q=3$. Both cases can be treated in complete
analogy to Example 2 above, and result in novel
sum rules
and combinatorial identities.

Recently we proved \cite{us} that exact trace formulae exist
for one-dimensional square wells with an arbitrary
number of potential steps inside.
Following the methods outlined above, our results can
be generalized immediately to obtain novel
combinatorial identities for necklaces with
more than two types of beads.

\section{Acknowledgments}
The authors acknowledge helpful comments and suggestions
by Rick Jensen.
Yu.D. and R.B. gratefully acknowledge financial support by
NSF grants PHY-9900730 and PHY-9984075; Yu.D.
by NSF grant PHY-9900746.

%
\pagebreak
.
\vskip2truecm
$$
\matrix{
 j &{\rm w}_j &w_j
   &\nu_j &n(w_j) &\alpha(w_j) &\beta(w_j) &\gamma(w_j) &\chi(w_j)\cr
 1 &\cal{LL}  &\cal{L}
   &2     &1      &1           &0          &2           &1        \cr
 2 &\cal{LR}  &\cal{LR}
   &1     &2      &0           &2          &3           &2        \cr
 3 &\cal{RR}  &\cal{R}
   &2     &1      &1           &0          &1           &2     \cr
 }
$$
 
\bigskip\noindent
{\bf Table~I:} List of the three cyclically non-equivalent binary
necklaces of length 2 together 
with their
primitives ($w$) and repetition indices ($\nu$). Some properties of the
primitives, such as their lengths ($n$), number of $\cal{R}$ or
$\cal{L}$ pairs ($\alpha$), number of transmissions ($\beta$), 
their weighted lengths ($\gamma$) 
and their phases ($\chi$) are also listed.
\par
\vskip2truecm
$$
\matrix{
 j &{\rm w}_j &w_j
   &\nu_j &n(w_j) &\alpha(w_j) &\beta(w_j) &\gamma(w_j) &\chi(w_j)\cr
 1 &\cal{LLLL}  &\cal{L}
   &4     &1      &1           &0          &2           &1        \cr
 2 &\cal{LLLR}  &\cal{LLLR}
   &1     &4      &2           &2          &7           &4        \cr
 3 &\cal{LLRR}  &\cal{LLRR}
   &1     &4      &2           &2          &6           &5        \cr
 4 &\cal{LRLR}  &\cal{LR}
   &2     &2      &0           &2          &3           &2        \cr
 5 &\cal{LRRR}  &\cal{LRRR}
   &1     &4      &2           &2          &5           &6        \cr
 6 &\cal{RRRR}  &\cal{R}
   &4     &1      &1           &0          &1           &2        \cr
 }
$$

\bigskip\noindent
{\bf Table~II:}
List of the six cyclically non-equivalent binary
necklaces of length 4. The meaning of the columns is the same
as in Table~I.
\end{document}